# The Wonders of Levitation

M. M. J. French

H. H. Wills Physics Laboratory, University of Bristol, Tyndall Avenue, Bristol. BS8 1TL. UK.

E-mail: mail@matthewfrench.net

**ABSTRACT**
I discuss some interesting classroom demonstrations of diamagnetism and how this effect can produce levitation. The possibilities for hands-on demonstrations of diamagnetic and superconducting levitation are discussed. To conclude I discuss some practical uses for levitation in daily life.

**INTRODUCTION**

We are perhaps all familiar with the wide range of experiments which can be used to demonstrate magnetism. Usually these show ferromagnetism or electromagnetism, but when it comes to demonstrations, their often overlooked cousin is diamagnetism. A diamagnetic material creates a magnetic field which opposes (at least to some extent) any external magnetic field the material is subject to. Diamagnetism is an effect arising from the change in the orbital velocity of the electrons around the nucleus of the atom in response to an applied magnetic field.

All materials show this diamagnetic response to an applied field, although, because this is a weak effect, it is overpowered in those materials which also exhibit ferromagnetism or paramagnetism. Common materials such as water, organic compounds (such as wood, plastics) and metals (copper, gold, mercury, bismuth) are all diamagnetic, but only very slightly. This ability to repel magnetic fields gives us the possibility of being able to levitate objects!

This has led to some interesting demonstrations such as the levitation of a live frog [1] and the diamagnetically stabilised levitation of a permanent magnet between human fingers [2]. These are spectacular demonstrations, but can only be achieved in the presence of an extremely large external magnetic field of around 10T (around 100,000 times greater than the Earth's magnetic field). Unfortunately, this prevents them being shown in the classroom, but images and videos can be found online which make exciting viewing.

**DIAMAGNETIC LEVITATION**

Diamagnetism can demonstrated in the classroom using permanent magnets and a small square of the best diamagnet at room temperature: pyrolytic graphite. This grows very slowly by a process called chemical vapour deposition, creating a highly ordered material where the carbon atoms form a layered hexagonal structure. Because of the low density of pyrolytic graphite a thin sheet will be repelled by a sufficiently strong neodymium magnet. A thick piece will be too heavy as the material above about a half of a millimetre does not contribute much to the lift. If the piece is thin enough, it will simply slide right off the side of a single magnet, and refuse to sit still on it. To get a piece of pyrolytic graphite to sit still above a magnet, we need to find a way to force it to the centre of a magnet.

This can be achieved by using a set of four magnets (creating a magnetic potential well). The pole of each of the magnets (where the field is strongest) forces the graphite to the centre. This effectively pins the graphite above the magnets. The diagram below shows the magnets coloured green and blue (with their North and South poles arranged as shown) and the pyrolytic graphite in grey in the centre. Note that the graphite sheet needs to be cleaved and cut slightly smaller than the size of a magnet.

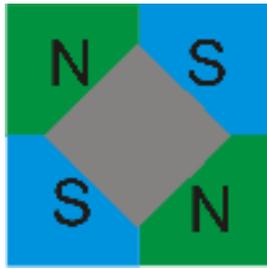

Figure 1: Arrangement of magnets and their polarity to levitate a piece of pyrolytic graphite.

This creates a stable room temperature levitation demonstration where the graphite is levitated about 1mm above the magnets. When pushed gently down, the graphite moves, but if the pressure is removed it levitates again. Although this demonstration is quite small (around 24mm x 24mm), it is available cheaply from [3]. Only 4 magnets are provided, but enough pyrolytic graphite is provided to make 8-10 units if additional magnets are purchased separately. This means around 5 - 8 can be used in each class allowing each student an opportunity to interact with it.

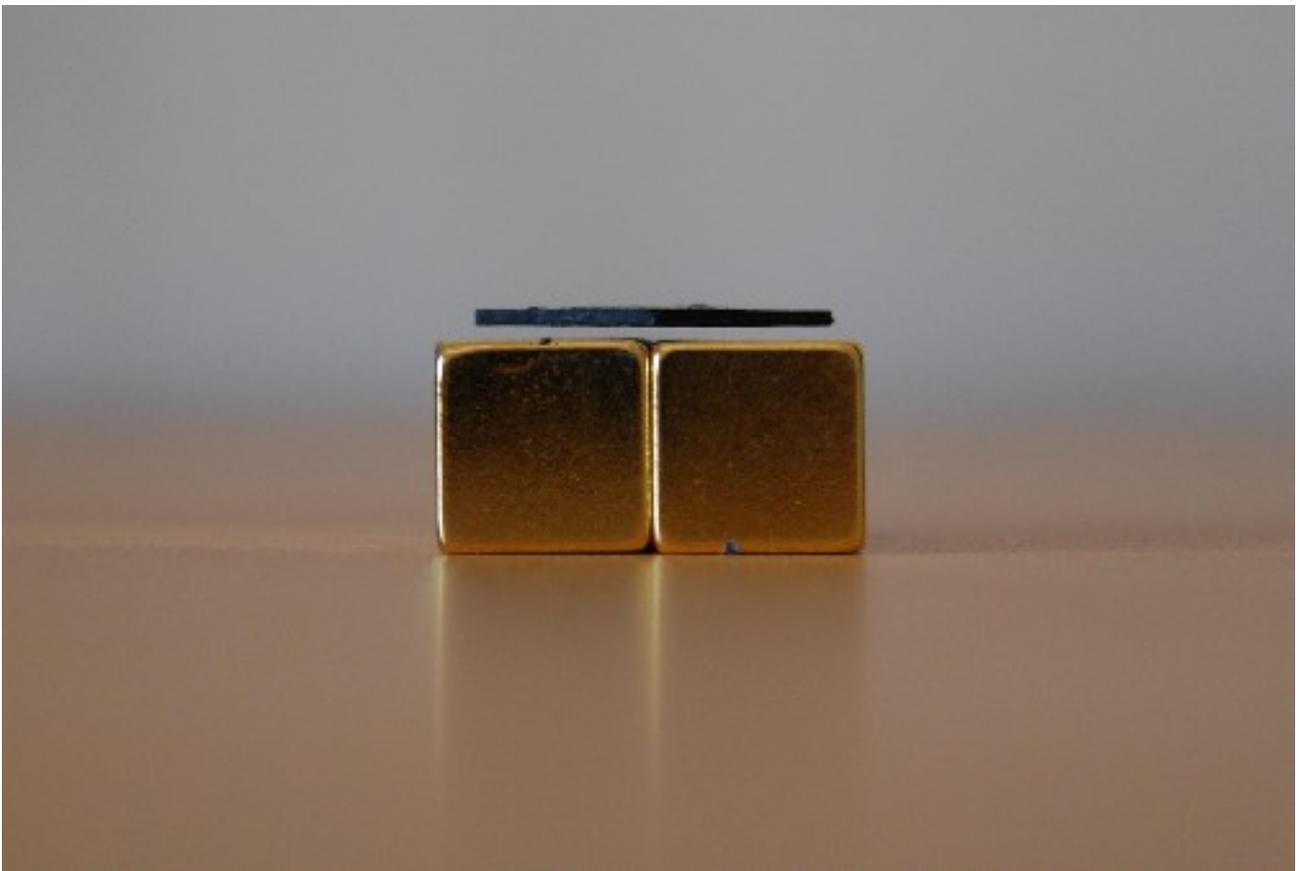

Figure 2: Photograph showing a piece of pyrolytic graphite levitating above permanent magnets.

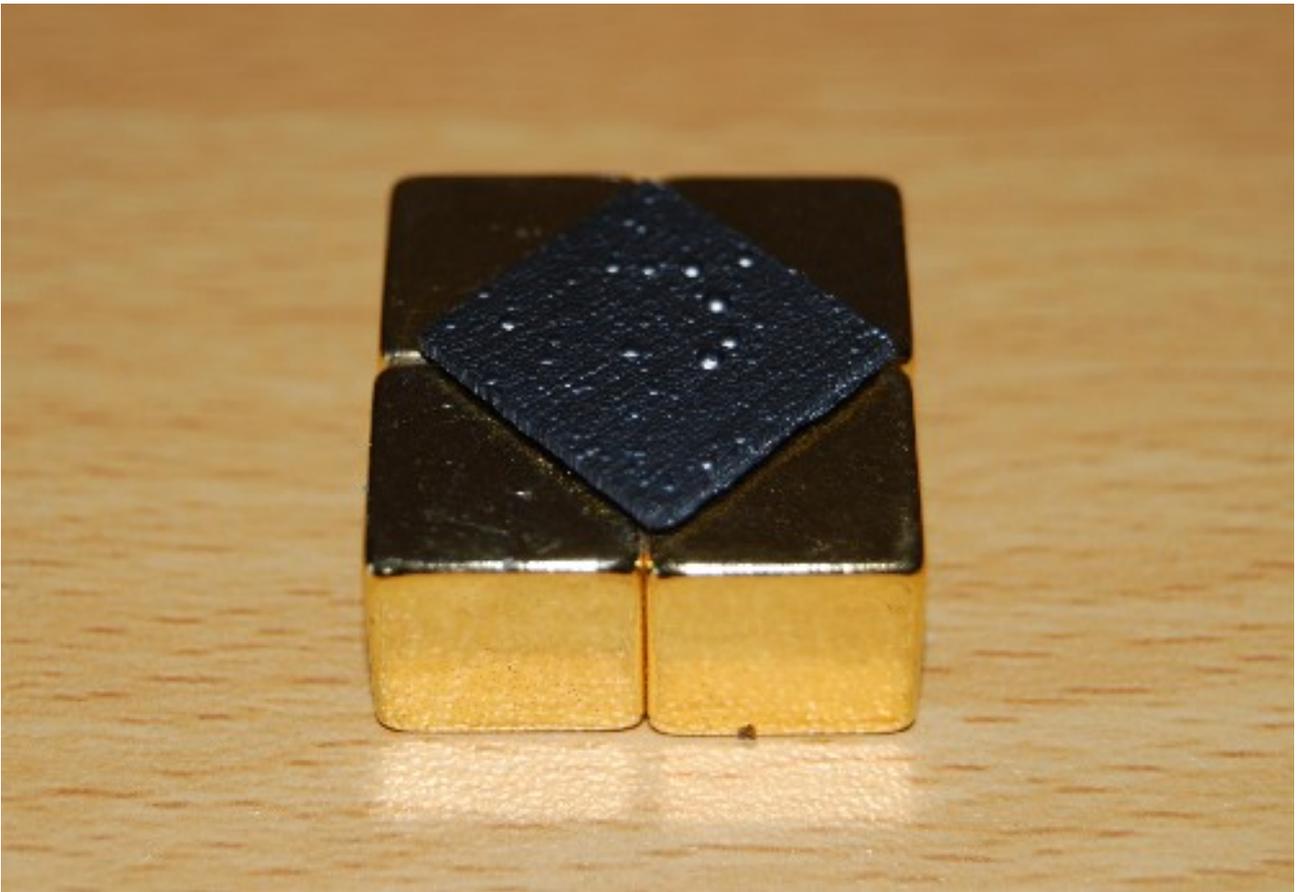
Figure 3: Photograph showing a piece of pyrolytic graphite levitating above permanent magnets.

**SUPERCONDUCTING LEVITATION**

An additional demonstration can be performed using superconductors, although this is only possible where a small amount of liquid nitrogen can be obtained to cool the superconductor below its superconducting transition temperature.

Superconductors are perfect diamagnets, meaning they will repel all external magnetic fields. A piece of superconducting material (Yttrium-Barium-Copper-Oxide), cooled to below its superconducting transition temperature (92K/-181°C) with liquid nitrogen (77K/-196°C) will levitate above a magnet. This is an example of Meissner effect - the magnetic field from the nearby permanent magnet can be expelled by the superconductor. This is because current flows in the surface of the material, generating a perfect mirror repulsive magnetic field.

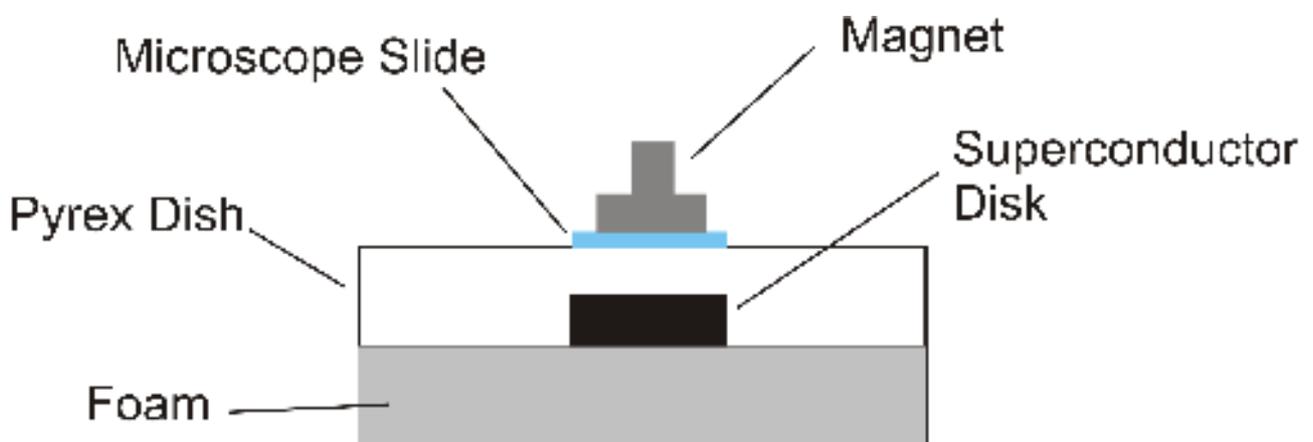
Figure 4: Diagram of apparatus needed for superconducting levitation demonstration.

This demonstration can be set up as shown in Figure 4 with a small Pyrex dish and some foam in the bottom. A disk of superconductor (available from [4]) is placed on the foam and a glass slide is balanced across the top of the dish with the magnet on top. The thickness of the foam should be such that there is around a 5-10mm gap between the superconductor and the magnet on the glass slide. Liquid nitrogen can then be carefully added to the Pyrex dish to cool the superconductor. Once cooled, the glass slide can be removed and the magnet levitates! The magnet can be spun, pushed down (it provides a much stronger 'resistance' to being pushed than the graphite discussed earlier) or even lifted up! There is another effect at work here which explains why the piece of superconductor is lifted up if the magnet is raised. If the superconductor is cooled in the presence of the magnetic field (or is pushed up close to a magnet) the magnetic field becomes 'pinned' into the superconductor by the presence of impurities in the superconductor.

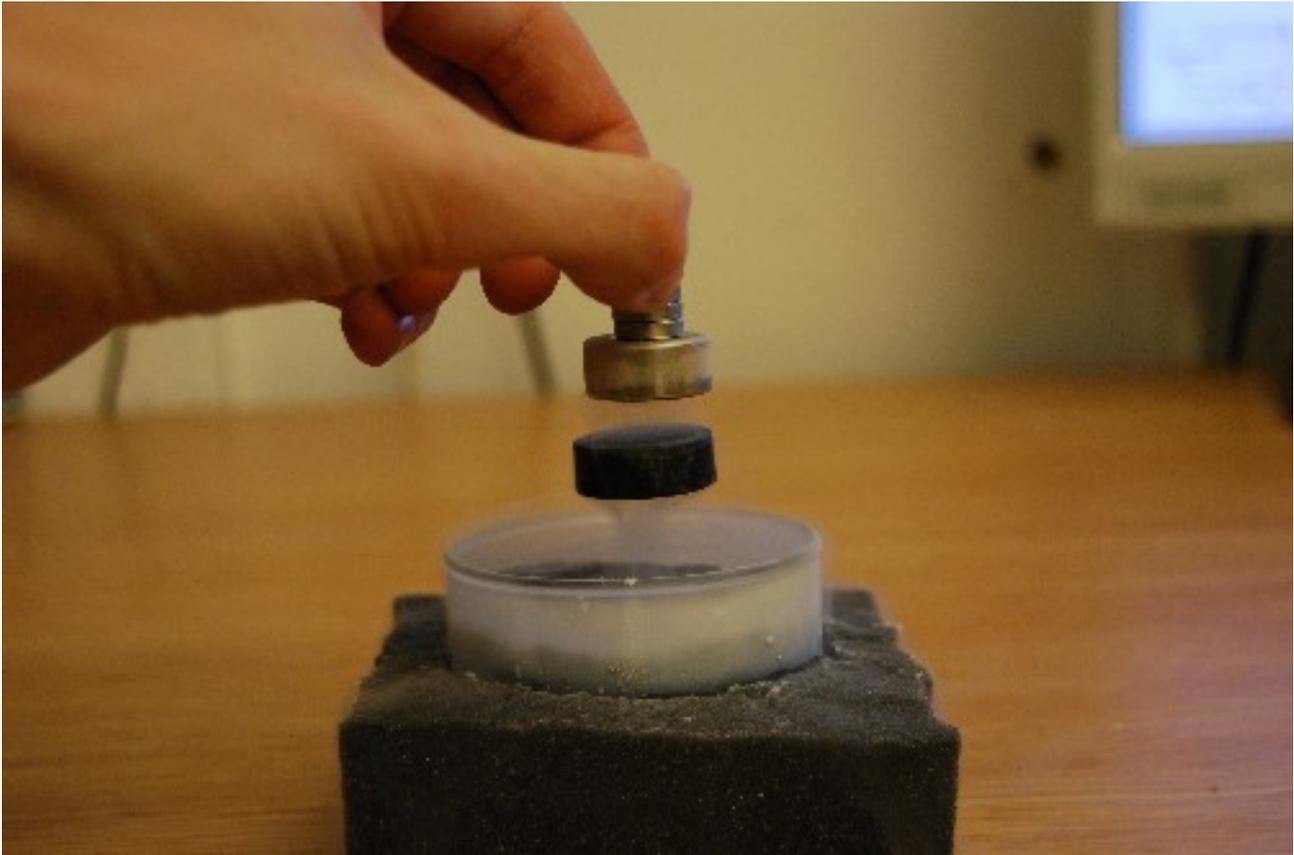

Figure 5: Superconducting levitation demonstration using a neodymium magnet (silver) and a superconductor levitation disk (black).

**MAGNETIC TRACK**

A further demonstration can be performed using a magnetic track: the superconductor is cooled in a vessel of liquid nitrogen and then placed onto the track and pushed down (to pin the magnetic field). It levitates and oscillates back and forth. This can also be adapted into an oval/circular track which the superconductor circles around.

The magnets in the track need to be set up to create a magnetic valley - so the track needs to be three magnets wide, usually with the magnets stuck to a piece of steel. Each magnet along the length of the track needs to have a top face of one pole and the opposing downward face of the other pole. Looking across the track three magnets are needed, the two edge magnets have, say north upwards and the middle magnet has south upwards (see Figure 6). On a linear track, the superconductor can be stopped and sent in the reverse direction at the ends by reversing the arrangement of the magnets so that the outer magnets are south upwards and the inner magnet is north upwards. If liquid nitrogen is not available, then videos of superconducting levitation can be found online.

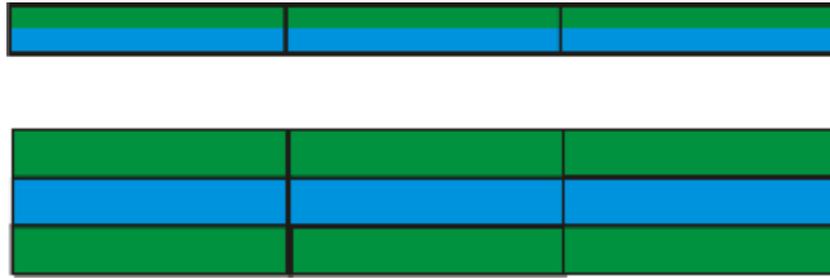

Figure 6: Configuration of magnet for a magnetic track. Green is north and blue is south.

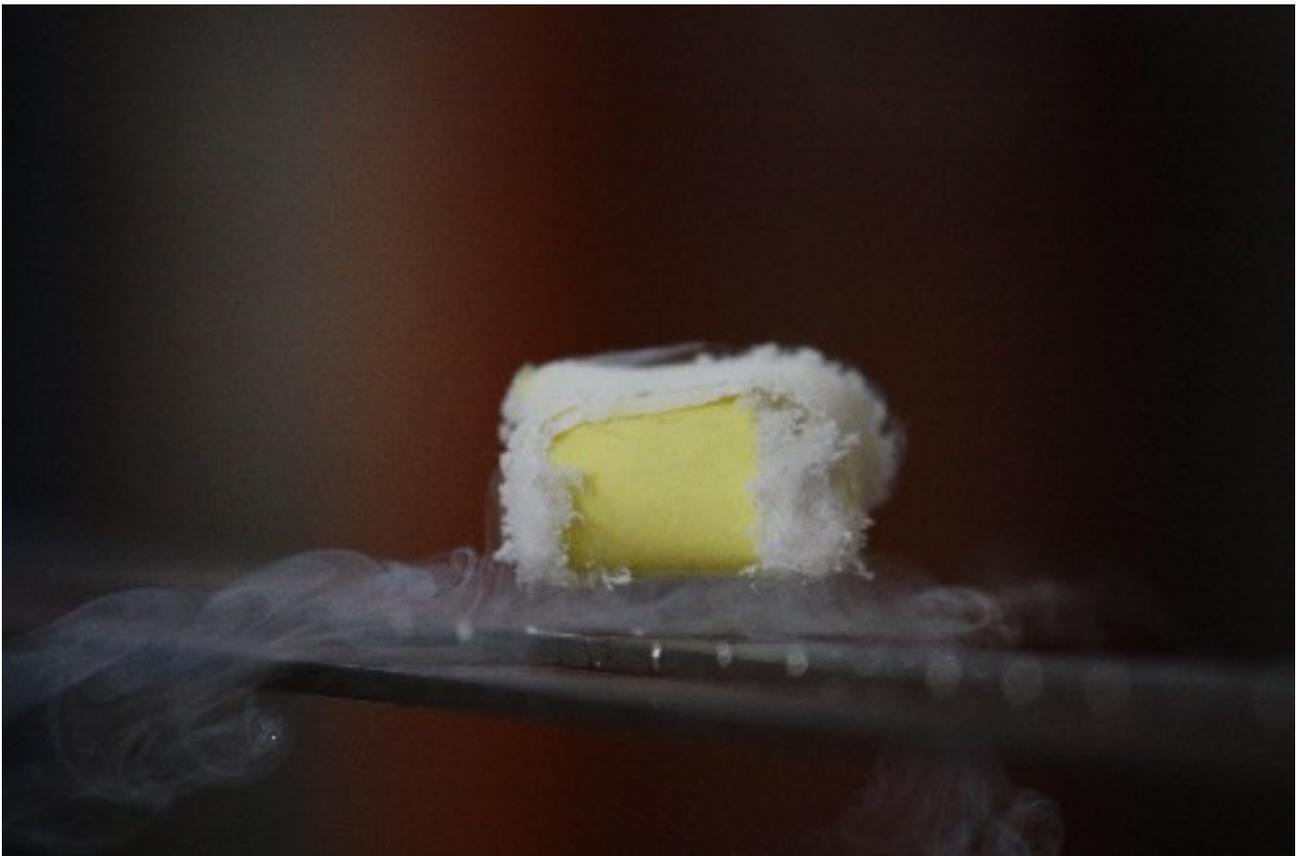

Figure 7: Photo of a piece of superconductor (wrapped in cotton wool and yellow sticky tape for thermal insulation) levitating above a magnetic track.

**USES**

Practical applications of levitation include frictionless bearings for use in industry and levitating trains. The fastest train in the world is currently the JR-Maglev in Yamanashi, Japan which levitates using the repulsive force generated when large currents flow through superconducting coils of wire. It has reached a top speed of 581 km/h (361 mph). Also in Japan a sumo wrestler weighing nearly 150kg has been levitated using a larger version of the superconducting levitation demonstration discussed earlier!

**ACKNOWLEDGEMENTS**

I thank Liam Malone and Jon Fletcher at the University of Bristol for helpful discussions and suggestions. I thank Chris Archer for the photograph used as figure 7 and also for help along with Liam Malone and Laurence French in creating the video enhancements to this article. This work was partially supported by outreach money provided by a portfolio grant from EPSRC.

**REFERENCES**

[1] http://www.hfml.ru.nl/levitate.html
[2] A. K. Geim, M. D. Simon, M. I. Boamfa, I. O. Heflinger, Nature 400 323, 1999
[3] https://www.scitoyscatalog.com
[4] http://www.can-superconductors.com